\documentclass[reprint,amsmath,amssymb,aps,prl,nofootinbib,superscriptaddressi,a4paper]{revtex4-2}
\usepackage[showcrop,top=0.5in, bottom=0.50in, left=0.6in, right=0.45in,marginparwidth=0in]{geometry}

\usepackage{graphicx}
\usepackage{color}
\usepackage{epstopdf}
\usepackage[usenames,dvipsnames]{xcolor}
\usepackage{bm}
\usepackage{paralist}
\usepackage{amsmath}
\usepackage{amsthm}
\usepackage{amssymb}
\usepackage{pstool}
\usepackage[percent]{overpic}
\usepackage{rotating}
\usepackage{lipsum}
\usepackage[normalem]{ulem}
\definecolor{Xred}{HTML}{B31B1B}

        \skip\footins 15pt
\usepackage{scalerel}
\usepackage{tikz}
\usetikzlibrary{svg.path}

\definecolor{orcidlogocol}{HTML}{A6CE39}
\tikzset{
  orcidlogo/.pic={
    \fill[orcidlogocol] svg{M256,128c0,70.7-57.3,128-128,128C57.3,256,0,198.7,0,128C0,57.3,57.3,0,128,0C198.7,0,256,57.3,256,128z};
    \fill[white] svg{M86.3,186.2H70.9V79.1h15.4v48.4V186.2z}
                 svg{M108.9,79.1h41.6c39.6,0,57,28.3,57,53.6c0,27.5-21.5,53.6-56.8,53.6h-41.8V79.1z M124.3,172.4h24.5c34.9,0,42.9-26.5,42.9-39.7c0-21.5-13.7-39.7-43.7-39.7h-23.7V172.4z}
                 svg{M88.7,56.8c0,5.5-4.5,10.1-10.1,10.1c-5.6,0-10.1-4.6-10.1-10.1c0-5.6,4.5-10.1,10.1-10.1C84.2,46.7,88.7,51.3,88.7,56.8z};
  }
}

\newcommand\orcidicon[1]{\href{https://orcid.org/#1}{\mbox{\scalerel*{
\begin{tikzpicture}[yscale=-1,transform shape]
\pic{orcidlogo};
\end{tikzpicture}
}{|}}}}

\newtheorem{theorem}{Theorem}
\newtheorem*{theorem*}{Theorem}
\newtheorem*{corollary*}{Corollary}
\newtheorem*{lemma*}{Lemma}
\newtheorem{lemma}{Lemma}

\newtheorem*{assumption*}{Assumption}
\newtheorem{assumption}{Assumption}
\newtheorem*{principle*}{Principle}

\usepackage[colorlinks=true,linkcolor=Magenta,citecolor=Magenta,breaklinks=true]{hyperref}

\def\ket#1{\mathinner{|{#1}\rangle}}

\begin{document}

\title{{\LARGE{ \bf A principle of information conservation for physical laws}}\\\mbox{}\vspace{-5pt} {\textnormal{\large{(Hidden information in quantum systems?)}}}}

\author{Nicolas G. Underwood\,\orcidicon{0000-0003-4803-2629}}
\date{\today}

\begin{abstract}
\hspace{-\parindent}%
A principle of information conservation is shown in abstract terms to rule out probabilistic physical laws, necessitating the existence of state trajectories.
It furthermore provides a geometric-thermodynamic mechanism for the appearance of probability distributions at the operational level, and thus may provide a dynamical explanation for Born's rule of quantum probabilities. 
This link between geometry and operational probabilities is argued to be a promising angle from which to study the potential for ``hidden information'' in quantum systems, and guide efforts in quantum foundations more generally.
\end{abstract}

\maketitle
\hspace{-\parindent}
In quantum theory, information is studied almost exclusively on an operational footing \cite{nielsen_chuang,Barrett_GPTs,H01,CAP11,MM11}.
If quantum systems do possess an underlying ``ontic'' state \cite{HS10}, however, there are many possible reasons why the information contained in this state may be operationally inaccessible.
The standard quantum formalism even serves to ``hide'' such information, so to speak.\footnote{To see this, note that 1) The PBR theorem tells us that each ontic state may correspond only to a single quantum state \cite{PBR,H13}. 2) Distinct ensembles of quantum systems may correspond to the same density matrix \cite{nielsen_chuang}.
Thus the standard formalism considers certain ``ontologically distinct'' ensembles as operationally equivalent, and so does not distinguish between them in the mathematics.}
In contrast, a realist\footnote{The field of quantum foundations is often described in terms of realism and operationalism, with textbook quantum mechanics arguably featuring elements of both. (The state trajectory being the realist part and the measurement theory being the operational part.) See Ref.~\cite{REALISM_VS_OPERATIONALISM} for a summary of the contrasting approaches and Refs.~\cite{BELLER,BV09} for historical perspectives.} perspective on information enables the consideration of such ``hidden information''.
While the term ``hidden'' appears to imply undetectability prima facie, theories with such hidden information have been shown, in certain circumstances, to have the potential to become empirically distinct \cite{UV15,UV16,SC20}, or leave observable signatures \cite{CV15,CV16}, leading to novel ways to test quantum theory \cite{AV07,AV10}.
Famously, a known unknown is more useful than an unknown unknown.

Rather than considering information in established quantum theories directly, the purpose of this paper is to demonstrate the role of realist information in physical theories abstractly, with as much generality, and with as few assumptions as reasonably possible\footnote{Cf.~several recent axiomatic or near axiomatic operational accounts of quantum mechanics \cite{H01,CAP11,MM11,Barrett_GPTs}.}, so as to establish a framework to help guide future investigations.
The primary result is a mathematical equivalence between information conservation and trajectories.
Specifically, physical laws conserve a straightforward classical notion of information [Eq.~\eqref{entropy}] if and only if they require the existence of deterministic state trajectories, which are geometrically constrained [Eq.~\eqref{divergence_free}]. (Probabilistic evolution is ruled out.)
Associated with this abstract information-based mechanics is a primitive information-based statistical mechanics, whereby trajectories spontaneously relax towards the geometrically defined distribution of maximum entropy, an instance of the second thermodynamic law.
This may create the appearance of indeterminism on an operational level.
As described below, for instance, it is straightforward to use the framework to devise models in which the Born rule of quantum probabilities arises thermodynamically, and therefore the reproduce the predictions of quantum theory.

Classical mechanics fits neatly into the framework, and indeed much of classical mechanics (e.g., symplectic geometry) may be shown to derive from information conservation as applied to canonical coordinates \cite{Uthesis}.
In contrast, both canonical quantum theory and de Broglie-Bohm pilot wave theory are only partial fits to the framework, and require qualification. 
Both are briefly commented upon below. 

To preserve accessibility to as broad an audience as possible, geometric mathematics is avoided. 
See Chp.~2 of Ref.~\cite{Uthesis} for differential geometric details.
The framework relies on the following two assumptions:
\begin{assumption}[State space]\label{assumption1}
The state of a system is completely described by $n$ real variables $x\in\mathbb{R}^n$, and these $n$ variables span the state space $\Omega$. If there exist boundaries, nontrivial topologies, or other complicating factors, it is assumed these can be accounted for in some reasonable manner so that their presence may be suppressed in the mathematics.
It is further assumed that distributions $\rho$ and $\mu$ may be integrated with respect to these coordinates, so that expressions like $\int_\Omega d^nx\rho(x)$ possess their elementary meaning. 
\end{assumption}
\begin{assumption}[Information measure]\label{assumption2}
Information regarding the state of a system at time $t$ is equivalent to the specification of a probability distribution upon the state space, $\rho(x,t)$.
The information content of $\rho(x,t)$ is quantified by the negative of the Jaynes entropy,
\begin{align}\label{entropy}
\mathfrak{Inf}[\rho(t)]:=-S_\text{Jaynes}(t):=\int_\Omega d^{n}x\, \rho(x,t)\log\frac{\rho(x,t)}{\mu(x)},
\end{align}
which is a carefully derived extension of the Shannon entropy to a continuous state space. (See p374, Ref.~\cite{Jaynesbook}.)
Distribution $\mu(x)$ may be equivalently understood as a density of states or as the distribution of maximum entropy.
\end{assumption}
\hspace{-\parindent}The Jaynes entropy \eqref{entropy} corrects for some well known deficiencies in so-called differential entropy $S_\text{diff}:=-\int d^nx\rho\log\rho$.
Primarily, an information measure should not depend upon the coordinates used to calculate it. 
Discrete entropy $S_\text{disc}:=-\sum_ip_i\log p_i$ is unaffected by a relabeling of states $i\longmapsto j(i)$, for instance. 
Yet $S_\text{diff}$ is not invariant under a change of coordinates, $x\longmapsto y(x)$.
Furthermore, Shannon's theorem establishing $S_\text{disc}$ as the correct information measure for a discrete space (theorem 2 in Ref.~\cite{Shannon}) does not follow over to $S_\text{diff}$.
The Jaynes entropy corrects for these shortcomings, and follows from $S_\text{disc}$ by assigning discretely labeled states a continuous coordinate and allowing these to become dense in the space $\frac{1}{n}\sum_{x_i\in\omega}1 \longrightarrow \int_\omega\mathrm{d}x\, \mu(x)$.
It is in this sense that $\mu(x)$ corresponds to a density of states.
Correspondingly, the proportion of total states contained in volume $\omega\subseteq\Omega$ is given by 
\begin{align}
N_\omega:=\int_\omega d^nx\mu(x)\leq1,
\end{align}
where the equality holds when region $\omega$ is equal to the entire space $\Omega$.
It is well known that the principle of indifference does not always carry over easily to continuous states (this is the content of Bertrand's paradox \cite{Bertrand}), and for this reason density $\mu(x)$ may or may not be uniform with respect to coordinates $x$.\footnote{For further discussion on this geometric connection see chapter 2 of Ref.\cite{Uthesis}, and for discussion of accounts of quantum theory based upon nonuniform geometries, see Ref.\cite{R15} and Refs.~therein.}
See Refs.~\cite{Uthesis,Jaynesbook} for derivations and further discussion on the Jaynes entropy. 

$\mathfrak{Inf}[\rho]$ intuitively quantifies the notion of information as follows.
Zero knowledge of a system's state means every state must be equally likely, i.e.~$\rho(x)=\mu(x)$, and corresponds to $\mathfrak{Inf}[\rho]=0$.
The opposite of this is perfect knowledge of the exact state of the system, i.e.~$\rho(x) \longrightarrow\delta^{(n)}(x-x_0)$ for some $x_0$, and corresponds to $\mathfrak{Inf}[\rho]\longrightarrow\infty$.
Between these two extrema, if the system is known to reside within some subregion of the state space $\omega\subset\Omega$, i.e.
\begin{align}\label{uniform_distribution}
\rho_{\omega}(x):=
\begin{cases}
\mu(x)/N_{\omega} &\text{ on }x\in \omega \\
0& \text{ on } x\in \omega_c
\end{cases},
\end{align}
where $\omega_c$ is the complement of $\omega$, then $\mathfrak{Inf}[\rho_{\omega}]=|\log N_\omega|$, in accord with Boltzmann's landmark expression.

On these assumptions impose information conservation:
\begin{principle*}[Information conservation]
A physical law is required to conserve the information content of any probability distribution $\rho$, so that $\mathfrak{Inf}[\rho(t)]=\mathfrak{Inf}[\rho(0)]$. 
\end{principle*}

\hspace{-\parindent}To explore the consequences of information conservation, introduce propagator $T(x,x',t)$, which is defined by its action on probability distribution $\rho$, 
\begin{align}\label{evolution}
\rho(x,t)=\int_{\omega'}d^nx'\,T(x,x',t)\rho(x',0),
\end{align}
for time variable $t\geq 0$.
This is a continuous analogue of the transition matrix of Markov Chains.
In the absence of information conservation, it may transform distributions $\rho$ in an arbitrary fashion, allowing for general, probabilistic physical laws. 
For instance, it may account for the possibility of a one-to-many probabilistic law if for a given $x',t$, distribution $T(x,x',t)=f(x)$ has non-singular support.
It also may account for the possibility of a many-to-one law if for a given $x,t$, distribution $T(x,x',t)=f(x')$ has non-singular support.
As a system definitely located in some state $x_0$ evolves to $\rho(x,t)=T(x,x_0,t)$, it must satisfy the basic requirements of a probability distribution, i.e.~$T(x,x',t)\geq0$ and $\int_\Omega d^nxT(x,x',t)=1$.
The content of theorem \ref{theorem1}, however, is that for every state $x'$ and time $t$, information conservation requires it to equal $T(x,x',t)=\delta^{(n)}(x-x')$ for a unique corresponding state $x$.
The resulting one-to-one mapping of states, $x',t\longmapsto x$, is taken to be the definition of a trajectory.

The following proofs are further facilitated by the introduction of the ``mask'' distribution $U_\omega(x,t)$, defined as
\begin{align}\label{mask}
U_{\omega'}(x,t):=\int_{\omega'}d^nx\, T(x,x',t)\mu(x').
\end{align}
Under propagation \eqref{evolution}, $T(x,x',t)$ transforms uniform distribution \eqref{uniform_distribution} as 
\begin{align}\label{evolution2}
\rho_{\omega'}(x',0)\longmapsto U_{\omega'}(x,t)/N_{\omega'}.
\end{align}
Accordingly, if the mask vanishes, $U_{\omega'}(x,t)=0$, then the physical law disallows any system initially in region $\omega'$ to be reside in state $x$ at time $t$.

\begin{lemma}[The usefulness of masks]\label{lemma1}
For any given $\omega'\subseteq\Omega$ and $t$, $U_{\omega'}(x,t)=\mu(x)$ on some region $x\in\omega\subseteq\Omega$, and $U_{\omega'}(x,t)=0$ on its complement $\omega_c$.
\begin{proof}
Transformation \eqref{evolution} is required to conserve $\mathfrak{Inf}[\rho]$ for any arbitrary initial distribution $\rho(x',0)$.
For a uniform distribution \eqref{uniform_distribution}, it is a short task to show that condition $\mathfrak{Inf}[\rho(t)]=\mathfrak{Inf}[\rho(0)]$ reduces to 
\begin{align}\label{condition2}
0=\int_\Omega d^nx\, U_{\omega'}(x,t)\log \frac{U_{\omega'}(x,t)}{\mu(x)},
\end{align}
which is required to hold for any region $\omega'$.
Take this region to be the entire space $\Omega$. 
As a uniform distribution on the entire space is uniquely low in information, this distribution must be conserved by the physical law.
By \eqref{evolution2} then, $U_\Omega(x,t)=\mu(x)$, which satisfies condition \eqref{condition2} as a result of the argument of the logarithm equaling unity for all $x$.
For the mask of a subregion, however, by definition \eqref{mask} $U_{\omega'}(x,t)\leq U_\Omega(x,t)$ for all $\omega'\subset \Omega$.
So the argument of the logarithm is $\leq 1$, with the result that the integrand in \eqref{condition2} is $\leq0$ for all $x$.
In order for the equality in condition \eqref{condition2} to hold then, the integrand must vanish for all $x$, so $U_{\omega'}(x,t)$ must equal either $\mu(x)$ or $0$ for all $x$, $\omega'$, and $t$.
\end{proof}
\end{lemma}
\begin{lemma}[No many-to-one]\label{lemma2}
No two distinct states $x_1'\neq x_2'$ may map to the same state $x$.
\begin{proof}
Suppose states $x_1'$ and $x_2'$ both have a possibility of being mapped to some state $x$.
By lemma \ref{lemma1} then, $U_{\omega_1'}(x,t)=U_{\omega_2'}(x,t)=\mu(x)$ for some $\omega_1'$ containing $x_1'$ and some $\omega_2'$ containing $x_2'$.
Lemma \ref{theorem1} will also apply to the union of the two regions, $U_{\omega_1'\cup\omega_2'}(x,t)=\mu(x)$.
If $x_1'$ and $x_2'$ are indeed two distinct states, however, then it will be possible to choose these regions to be non-overlapping, $\omega_1'\cap\omega_2'=\{\}$.
But by definition \eqref{mask}, $U_{\omega_1'\cup\omega_2'}(x,t)=U_{\omega_1'}(x,t)+U_{\omega_2'}(x,t)=2\mu(x)$, which is a contradiction.
\end{proof}
\end{lemma}
\begin{lemma}[No one-to-many]\label{lemma3}
No single state $x'$ may map to two distinct states $x_1\neq x_2$.
\begin{proof}
Lemma \ref{lemma2} prevents many-to-one maps, yet the physical law may still map states in a one-to-many probabilistic fashion, onto subsets of the state space $x'\longmapsto \omega|_{x',t}\subset\Omega$. 
By lemma \ref{lemma2}, however, these subsets cannot overlap, $\omega|_{x'_1,t}\cap \omega|_{x'_2,t}=\{\}$ for any $x'_1\neq x'_2$.
Furthermore, as zero information distribution $\rho(x,t)=\mu(x)$ is conserved, and by definition $\mu(x)$ spans $\Omega$, it is clear the physical law is onto. 
Hence the inverse map $\omega|_{x',t}\longmapsto x'$ is well defined, and so then is the inverse propagator $T^{-1}(x,x',t)$.
Lemmas \ref{lemma1} and \ref{lemma2} may then be repeated with the propagator replaced with its inverse, meaning the proof of no many-to-one becomes a proof of no one-to-many.
\end{proof}
\end{lemma}
\begin{theorem}[Information conservation implies trajectories]\label{theorem1}
A physical law that conserves information is a one-to-one and onto mapping on the state space. 
\begin{proof}
One-to-one-ness follows from lemma \ref{lemma2} and lemma \ref{lemma3}. Onto-ness follows from conservation of zero information distribution $\rho(x,t)=\mu(x)$.
\end{proof}
\end{theorem}
\hspace{-\parindent}As the presence of trajectories has now been established, the propagator may be formally written
\begin{align}
T(x,x',t)=\delta^{(n)}[x-x(x',t)],
\end{align}
where the trajectory beginning in state $x'$ is implied by the notation $x(x',t)$.
Of course unconstrained trajectories do not guarantee information conservation alone; it is a simple task to devise trajectories that violate information conservation (see Ref.~\cite{Uthesis} for pedagogical examples). 
Trajectories are constrained by (version 2 of) Liouville's theorem, which is traditionally stated in two different forms.
The first version follows as a simple corollary of lemma \ref{lemma1} while the second presumes the presence and differentiability of trajectories, requiring one further assumption. 
\begin{assumption}[Differentiability of trajectories]\label{assumption3}
The physical law implied by the trajectory of theorem \ref{theorem1} and notation $x(x',t)$, is sufficiently well behaved for derivatives to be taken with respect to coordinates $x$, $t$ so that expressions like $\dot{x}$ and $\nabla\cdot x$ possess their elementary meaning.
\end{assumption}

\begin{theorem}[Generalized Liouville's theorem]\label{theorem2}
Version~1) State space regions are mapped to regions of the same volume, as measured in number of states.
Version~2) Along trajectories $d/dt(\rho/\mu)$ is conserved, which is equivalent to condition
\begin{align}\label{divergence_free}
\nabla\cdot(\mu\dot{x})=0,
\end{align}
incompressibility of law $\dot{x}$ with respect to the state density.
\begin{proof}[Proof of version 1]
By lemma \ref{lemma1}, a uniform distribution on some region $\omega'$ is mapped to a uniform distribution on a new region $\omega$, equal to $\rho(x,t)=\mu(x)/N_{\omega'}$. For this to be normalized, it must be the case that $N_{\omega}=N_{\omega'}$, i.e.~the number of states in each region is the same. Equivalently, the Jacobian for the evolution is $\det\mathbf{J}=m(x)/m(x')$.
\end{proof}
\begin{proof}[Proof of version 2]
Assumption \ref{assumption3} means probability distributions $\rho$ follow the continuity equation $\partial\rho/\partial t+\nabla \cdot(\rho\dot{x})=0$.
Any continuity equation may be written $\rho^{-1}d\rho/dt=-\nabla\cdot\dot{x}$, where $d\rho/dt:=\partial/\partial t+\dot{x}\cdot\nabla$ is the material derivative along a trajectory. 
It follows that the ratio of any two distributions is conserved along any trajectory, $d/dt(\rho_1/\rho_2)=0$.
In the presence of information conservation, since zero information corresponds to $\rho(x,t)=\mu(x)$, it is the case that $d/dt(\rho/\mu)=0$.
Equivalently, the condition may be expressed $\nabla\cdot(\mu\dot{x})=0$.
\end{proof}
\end{theorem}
\hspace{-\parindent}The reverse implication to theorems \ref{theorem1} and \ref{theorem2}, i.e.~that information conservation follows from trajectories satisfying condition \eqref{divergence_free}, may be found through considering evolution $x'\longmapsto x(x',t)$ as a coordinate transform on Eq.~\eqref{entropy}, and invoking theorem \ref{theorem2} (see Ref.~\cite{Uthesis} for particulars). 
However in non-trivial, practical circumstances, information is lost on a \emph{de facto} basis, and spontaneous thermodynamic relaxation takes place towards the distribution of zero information, $\mu(x)$.
This is for precisely the same reason as entropy tends to rise in classical mechanics, and as such encompasses the classical tension between time-reversible dynamics and time-irreversible thermodynamics.  
Exact information is conserved, but information is lost on a \textit{de facto} basis through empirical inaccessibility of increasing microstructure in $\rho$ (see Ref.~\cite{Daviesbook}).
Intuitively, Josiah Gibbs' classical dye mixing analogy \cite{Gibbsbook}, which was motivated by the incompressibility of the flow, still works perfectly with the proviso that the incompressibility is with respect to the state density, not the coordinates.
The following theorem is only a slight adaptation of the classic textbook proof of entropy rise originally due to Ludwig Boltzmann, corrected by Paul Ehrenfest and Tatyana Afanasyeva, and adapted to quantum theory by Antony Valentini \cite{Daviesbook,AV91a}. 
\begin{theorem}[Thermodynamic relaxation (H-theorem)]\label{theorem3}
In any non-trivial, practical circumstance, information is lost and entropy rises.
\begin{proof}
Discretize the state space into small cells (so-called coarse graining) that reflect the limitations of experiment.
Presume that an experimenter attempting to calculate the change in entropy may only access distributional averages over these cells, denoted with a bar e.g., $\bar{\rho}=\int_{\delta\omega}d^nx\rho$.
Assume the volume of these cells, $\delta\omega$, to be small enough to capture the initial distribution $\rho(0)$ and density of states $\mu$, so that to a good approximation $\bar{\rho}(0)=\rho(0)$ and $\bar{\mu}=\mu$ in each cell, but not small enough to capture the microstructure in $\rho$, so that $\bar{\rho}\neq\rho$.
Recall that the exact entropy is conserved. Hence $\mathfrak{Inf}[\bar{\rho}(0)]=\mathfrak{Inf}[\rho(0)]=\mathfrak{Inf}[\rho(t)]$. 
The change in entropy may then be calculated as
\begin{align}
\Delta\mathfrak{Inf}[\bar{\rho}]
:&=\mathfrak{Inf}[\bar{\rho}(t)]-\mathfrak{Inf}[\bar{\rho}(0)]
=\mathfrak{Inf}[\bar{\rho}(t)]-\mathfrak{Inf}[\rho(t)]\nonumber
\\
&=\int_\Omega d^nx\left(\bar\rho\log\bar\rho - \bar\rho\log\mu - \rho\log\rho + \rho\log\mu\right).
\end{align}
Since $\log\mu=\log\bar\mu$, the second and fourth terms in the integrand cancel. 
Similarly, the first term may be replaced with $\rho\log\bar\rho$.
Hence, the total integrand reduces to $\rho\log(\bar\rho/\rho)$.
To this, add term $0=\int_\Omega d^nx\left(\rho-\bar\rho\right)$, so that
\begin{align}
\Delta\mathfrak{Inf}[\bar{\rho}]&=\int_\Omega d^nx\left(\rho\log\frac{\bar{\rho}}{\rho}+\rho-\bar\rho\right)\leq0.
\end{align}
The final inequality is found as function $a\log(b/a)+a-b$ is negative for $a\geq0$, $b\geq0$, $a\neq b$, and is equal to zero for $a\geq0$, $b\geq0$, $a=b$. 
\end{proof}
\end{theorem}
\hspace{-\parindent}\textit{\textbf{In summary,}} theorem \ref{theorem1} states that information conservation implies trajectories. Theorem \ref{theorem2} states these trajectories must be divergence free with respect to state density $\mu$, and results in the reverse implication. 
Thus a mathematical equivalence is found.
Information, as quantified by Eq.~\eqref{entropy}, is conserved if and only if there exist trajectories that are constrained by the geometry of the space according to Eq.~\eqref{divergence_free}.
This duality between information and trajectories is complemented by the second thermodynamic law in the guise of theorem \ref{theorem3}, which states that in practical circumstances information is lost on a \emph{de facto} basis; spontaneous thermodynamic relaxation takes place toward a distribution of zero information (maximum entropy) equal to the density of states $\mu$.
Hence, the thermodynamic properties of a physical system, give insight into its mechanics and state space geometry, and vice versa. 
This information framework may be used as a basis to derive much of classical mechanics \cite{Uthesis}, however it sits a little more uncomfortably with quantum theory.


The canonical trajectory of quantum theory is that of the quantum state $\ket{\psi}$ upon the (projective) Hilbert space $\mathcal{P}\subset\mathcal{H}$. 
The corresponding dual information measure is not the von Neumann entropy, but instead one that integrates a probability density $\rho(\ket{\psi})$ directly over $\mathcal{P}$ according to Eq.~\eqref{entropy} in some set of coordinates.
Assuming the Hilbert space admits a discrete energy basis $\ket{\psi}=\sum_n c_n\ket{E_n}$, Schr\"{o}dinger evolution $\ket{\psi}\longmapsto e^{-iHt/\hbar}\ket{\psi}$ produces a divergence free flow in the real and imaginary parts of $c_n$.  
Hence, there is a conserved information measure in these coordinates with a uniform $\mu$.
The relationship between this measure and the von Neumann entropy, which is discrete rather than continuous, is not immediately clear.
In particular it would be informative to know whether there is an equivalent of theorem \ref{theorem1} for the von Neumann entropy, so that its conservation might also guarantee trajectories.\\
\mbox{}\hspace{\parindent} A second place trajectories feature in quantum theory is in the de Broglie-Bohm pilot wave formulation.
Though undiscovered for many decades, it is now widely recognized that de Broglie's deterministic trajectories have the intriguing property of ``quantum relaxation'', meaning, modulo caveats, ensembles of such trajectories that are initially distributed arbitrarily, spontaneously relax towards the Born distribution, $\rho=|\psi|^2$ \cite{AV91a,VW05,CS10,Uthesis}. 
(See Chp.~2, Ref.~\cite{Uthesis} for numerous illustrative examples.)
This raises the possibility that Born's probability rule could have a dynamical origin, and that nonequilibrium violations of it may have left observable signatures or may even still be empirically accessible \cite{AV10,UV15,UV16}. 
For the present purposes, this relaxation property may be understood as an instance of theorem \ref{theorem3}, and it is straightforward to use the information framework described to construct model theories that ``quantum relax''. 
All that is required is for information conservation to be imposed upon a density of states $\mu=|\psi|^2$.
Then theorem \ref{theorem1} guarantees the existence of trajectories, and it is a simple process to find an equation of motion consistent with theorem \ref{theorem2} \cite{Uthesis}.
By theorem \ref{theorem3}, such trajectories would then be expected to relax to (or at least towards) $|\psi|^2$.
The proviso is that this construction involves a time dependent density of states, $\mu=|\psi(x,t)|^2$, so a coordinate in the state space is not sufficient to fully determine evolution. 
In other words, de Broglie-Bohm evolution presumes prior knowledge of the quantum state; coordinates $x$ do not exhaust the true state space.
This is hardly surprising as the process may be likened to going from a classical theory on a phase space $\{x\}=\{q,p\}$, to a quantum theory with a state space density $\mu(q)$ or $\mu(p)$, and so halving the number of state space parameters. 
It remains to be seen whether a unified description may be given, i.e.~one that fits fully into the framework, thus treating quantum and sub-quantum information on the same footing. 
However such a description could certainly lead to new insights into quantum theory in its many forms. \\
\mbox{}\hspace{\parindent}It is hoped that the central result of this work, namely the mathematical equivalence of information conservation and geometrically constrained trajectories, may be of broad appeal. 
It suggests, quite generally, that we should expect trajectories on the same level as we expect information, and that to theorize a modification or absence of one requires a necessary alteration of the other.
Information conservation and the geometrical connection to operational probability distributions appears a promising route towards an axiomatic realist account of quantum theory, and this would undoubtedly be valuable to the field.

\bibliographystyle{modified-hunsrt}

\end{document}